# Simulating Student Success in the Age of GenAI: A Kantian-Axiomatic Perspective

Seyma Yaman Kayadibi

Victoria University

seyma.yamankayadibi@live.vu.edu.au

## Abstract

This study reinterprets a Monte Carlo simulation of students' perceived success with generative AI (GenAI) through a Kantian-axiomatic lens. Building on prior work, theme-level survey statistics, Ease of Use & Learnability, System Efficiency & Learning Burden and Perceived Complexity & Integration, from a representative dataset, are used to generate 10,000 synthetic scores per theme on the [1,5] Likert scale. The simulated outputs are evaluated against the axioms of dense linear order without endpoints (DLO), irreflexivity, transitivity, total comparability (connectedness), no endpoints (no greatest and no least; A4-A5), and density (A6). At the data level, the basic ordering axioms (A1-A3) are satisfied, whereas no-endpoints (A4-A5) and density (A6) fail as expected: Likert clipping introduces minimum and maximum observed values, and a finite, discretized sample need not contain a value strictly between any two distinct scores. These patterns are read not as methodological defects but as markers of an epistemological boundary. Following Kant and Friedman, the findings suggest that what simulations capture, finite, quantized observations, cannot instantiate the ideal properties of an unbounded, dense continuum; such properties belong to constructive intuition rather than to finite sampling alone. A complementary visualization contrasts the empirical histogram with a sine-curve proxy to clarify this divide. The contribution is therefore interpretive rather than data-expansive: it reframes an existing simulation as a probe of the synthetic a priori structure underlying students' perceptions, showing how formal order-theoretic coherence coexists with principled failures of endpoint-freeness and density in finite empirical models.

## 1. Introduction

The rapid development of generative artificial intelligence (GenAI) systems, such as ChatGPT, has intensified interest in their pedagogical implications in higher education. As these tools become embedded in academic workflows, questions arise not only about practical utility but also about how students perceive, evaluate, and cognitively engage with them. In a prior study, Kayadibi (2025) introduced a hybrid methodological framework that combined a systematic literature review with simulation-based modeling to examine links between students' perceptions of GenAI and perceived academic success. Through a PRISMA-guided synthesis of nineteen empirical studies, six datasets reporting item-level means and standard deviations were identified. From these, one dataset was selected to simulate 10,000 synthetic student scores via a Monte Carlo approach with inverse-variance weighting, yielding a composite "Success Score" organized around three themes: Ease of Use & Learnability, System Efficiency & Learning Burden, and Perceived Complexity & Integration. The initial simulation indicated that usability- and functionality-related dimensions, rather than trust or general attitudes, were the strongest predictors of perceived success. While these findings offered a statistically grounded, perception-driven account of students' responses to GenAI, they left open a deeper theoretical question: to what extent does the simulation reflect the cognitive architecture through which such perceptions are structured?

The present paper addresses this question by embedding the earlier simulation within a Kantian-axiomatic framework. Drawing on Kant's theory of cognition, the three themes are interpreted not merely as empirical constructs but as epistemic conditions akin to Kant's a priori forms, e.g., space and time, that organize experience and make perception possible. Within this framework, the model is evaluated against the axioms of dense linear order (DLO), irreflexivity, transitivity, total comparability (connectedness), no endpoints (no greatest and no least element), and density, as classically associated with modern formulations of continuity in Hilbert's axiomatization and analyzed by Friedman (1992, pp. 62–64). Rather than treating simulation purely as a predictive statistical tool, the study asks whether such models can be read as cognitive approximations of Kantian perception: finite, discrete representations that strive, yet necessarily fail, to capture the continuous character of intuition. This shift repositions the simulation as a philosophical object: not only a methodological device but also a site of epistemological inquiry.

**Research Question.** To what extent does a Monte Carlo-based simulation of students' perceptions of GenAI conform to the axioms of dense linear order, and what does this reveal about the epistemological limits of empirical modeling in educational contexts?

**Study Objective.** To examine the internal logical structure of a Monte Carlo-based simulation of GenAI-related student success by testing its adherence to the axioms of dense linear order, and to reinterpret the simulation through a Kantian lens so as to delineate the boundary between statistical representation and cognitive synthesis.

## 2. Theoretical Integration of Method and Epistemology

To examine whether the educational themes extracted from empirical studies possess internal logical consistency, simulation-based modeling is integrated with a Kantian formal analysis. The working premise is that students' perceptions of GenAI systems are shaped not only by observable variables but also by deeper cognitive schemata akin to Kant's a priori forms. A Monte Carlo simulation is used to approximate how learners evaluate system features such as usability and integration, while Kant's transcendental philosophy provides the interpretive scaffolding for assessing whether those evaluations conform to the structural logic of cognition. In this sense, simulation is treated not merely as a statistical technique but as an epistemic instrument for probing the architecture of perception. The technical design is detailed in the next section.

## 3. Methodology

This study builds on Kayadibi (2025), which combined a PRISMA-guided review with Monte Carlo modeling to analyze students' perceptions of GenAI in education. The review synthesized nineteen empirical studies and identified six datasets reporting item-level means and standard deviations. From these six, one dataset (Veras et al., 2024) was selected for focused simulation because it provided sufficiently granular statistics to parameterize the three themes employed here: Ease of Use & Learnability, System Efficiency & Learning Burden, and Perceived Complexity & Integration. The present work extends that empirical base by applying a Kantian axiomatic analysis to the simulated results to test whether the generated patterns align with logical structures associated with human cognition.

### 3.1 Simulation Framework

A Monte Carlo simulation was conducted using theme-level statistics extracted from six empirical survey studies. Among these, one dataset was selected for simulation due to its comprehensive reporting of item-level means and standard deviations. This dataset enabled the generation of continuous, theme-specific distributions suitable for Monte Carlo modeling and subsequent axiomatic analysis. Based on the reported metrics, synthetic data was generated to simulate perceived academic success across three cognitive themes:

- Ease of Use & Learnability ($\mu = 4.1169$, $\sigma = 0.2709$)

•System Efficiency & Learning Burden (µ = 4.1240, σ = 0.0910)

•Perceived Complexity & Integration (µ = 3.7100, σ = 0.2160)

**Assumptions.** Each theme's scores are modeled as independent normal variables. Simulated values are clipped to the [1, 5] Likert range and rounded to four decimals to match the survey instrument. The implications of clipping, introducing endpoints, and discretization, are addressed in the axiomatic analysis. A fixed random seed is used for reproducibility. This simulated dataset serves as the empirical basis for the subsequent tests of dense linear order.

### 3.1.1 Python Code: Simulating Theme-Based Scores

The following code was used to generate the synthetic dataset:

```python
import numpy as np, pandas as pd

np.random.seed(42)
themes = {
    "Ease of Use & Learnability": {"mean": 4.1169, "std": 0.2709},
    "System Efficiency & Learning Burden": {"mean": 4.1240, "std": 0.0910},
    "Perceived Complexity & Integration": {"mean": 3.7100, "std": 0.2160}
}
num_students = 10000

df = pd.DataFrame({
    theme: np.clip(np.random.normal(loc=val["mean"], scale=val["std"],
size=num_students), 1, 5).round(4)
    for theme, val in themes.items()
})

pd.set_option('display.max_columns', None)
print(df.head(10))
```

*Figure 1. Simulated success scores for 10,000 students across three themes*

**Table 1. Sample of Simulated Theme-Based Success Scores (n = 10)**

Excerpted output from the Python-based Monte Carlo simulation described in Section 3.1.1. These 10 values are drawn from the 10,000-score synthetic dataset generated for each

| ID | Ease of Use & Learnability | System Efficiency & Burden | Perceived Complexity & Integration |
|---|---|---|---|
| 0 | 4.2515 | 4.0623 | 3.7852 |
| 1 | 4.0794 | 4.0962 | 3.7712 |
| 2 | 4.2924 | 4.0696 | 3.5077 |
| 3 | 4.5295 | 4.1340 | 3.8352 |
| 4 | 4.0535 | 4.2329 | 3.3881 |
| 5 | 4.0535 | 4.0538 | 3.5687 |
| 6 | 4.5447 | 4.2151 | 3.2782 |
| 7 | 4.3248 | 4.0529 | 4.0467 |
| 8 | 3.9897 | 4.0469 | 3.6600 |
| 9 | 4.2639 | 4.1985 | 4.1780 |

theme, and serve as illustrative samples for the subsequent axiomatic analysis.

*Table 1. The sample of 10 scores shown in Table 1 was selected for manual axiomatic evaluation*

**Ethical Considerations**

This study did not involve human participants or the collection of personal data. All analyses were conducted using published literature and simulation methods. Therefore, approval from a human research ethics committee was not required.

**4. Axiomatic Consistency Analysis**

**4.1 Theoretical Basis**

Following Friedman (1992, p. 62), the strict order "<" on X is governed by the theory of dense linear order without endpoints. The axioms are:

1. **Irreflexivity:** $\forall a \in X \ \neg(a < a)$.
2. **Transitivity (standard form):** $\forall a, b, c \in X \ ((a < b \land b < c) \Rightarrow a < c)$.
3. **Total comparability (connectedness):** $\forall a, b \in X \ ((a < b) \lor (b < a) \lor (a = b))$.
4. **No greatest element:** $\forall a \in X \ \exists b \in X \ (a < b)$.

5. **No least element:** $\forall a \in X \, \exists b \in X \, (b < a)$.
6. **Density:** $\forall a, b \in X \, (a < b \Rightarrow \exists c \in X \, (a < c < b))$.

## 4.2 Theme-by-Theme Results

### 4.2.1 Theme 1: Ease of Use & Learnability— Axiomatic Analysis

**Axiom 1: Irreflexivity ($\neg a < a$)**

Definition. No element is less than itself.

Mathematical Statement. $\forall a \in X, \neg(a < a)$.

Test. Each score was compared against itself (e.g., 4.2515<4.2515 → False).

Conclusion. Passed.

**Axiom 2: Transitivity**

Definition. Order relations must be preserved through intermediate comparisons.

Mathematical Statement. $\forall a, b, c \in X, (a < b \wedge b < c) \Rightarrow a < c$.

Test. $x_5 = 4.0535 < 4.0794 = x_2 < 4.2515 = x_0 \Rightarrow x_5 < x_0; x_2 = 4.0794 < 4.2924 = x_3 < 4.5295 = x_4 \Rightarrow x_2 < x_4$.

Conclusion. Passed.

**Axiom 3: Total comparability (connectedness)**

Definition. Every pair of elements is comparable: a<b, b<a, or a=b.

Mathematical Statement. $\forall a, b \in X, (a < b) \vee (b < a) \vee (a = b)$.

Test. Sample pairs, 4.2515 vs 4.0794; 4.2515 vs 4.2924; 4.2515 vs 4.5295, were all comparable.

Conclusion. Passed.

**Axiom 4: No greatest element**

Definition. For every a, there exists some b>a.

Mathematical Statement. $\forall a \in X, \exists b \in X\ (a < b)$.

Data-level test (sample of 10). A maximum exists: 4.5447. For a=4.5447, there is no $b \in X$ with a<b.

Conclusion. Failed (by construction on a finite subset).

**Axiom 5: No least element**

Definition. For every a, there exists some b<a.

Mathematical Statement. $\forall a \in X, \exists b \in X\ (b < a)$.

Data-level test, sample of 10. A minimum exists: 3.9897. For a=3.9897, there is no $b \in X$ with b<a.

Conclusion. Failed (by construction on a finite subset).

**Axiom 6: Density**

Definition. For any two distinct elements a<b, a third element must exist strictly between them.

Mathematical Statement. $\forall a, b \in X,\ \big(a < b \Rightarrow \exists c \in X\ (a < c < b)\big)$.

Test. Adjacent distinct pairs in the sorted set, 4.0535<4.0794, 4.2515<4.2639, have no value of X strictly between them.

Conclusion. Failed. A finite, discretized sample cannot be dense.

**4.2.2 Theme 2: System Efficiency & Learning Burden — Axiomatic Analysis**

Let

$$X = \{4.0623,\ 4.0962,\ 4.0696,\ 4.1340,\ 4.2329,\ 4.0538,\ 4.2151,\ 4.0529,\ 4.0469,\ 4.1985\}.$$

Sorted (unique):
4.0469<4.0529<4.0538<4.0623<4.0696<4.0962<4.1340<4.1985<4.2151<4.2329.

**Axiom 1: Irreflexivity** $(\neg a < a)$

Statement. $\forall a \in X \ \neg(a < a)$.

Test. Each value compared with itself (e.g., 4.0623<4.0623) evaluated False.

Conclusion. Passed.

**Axiom 2: Transitivity**

Statement. $\forall a, b, c \in X \ \big((a < b \land b < c) \Rightarrow a < c\big)$.

Test. 4.0469<4.0538<4.0623 $\Rightarrow$ 4.0469<4.0623; 4.0962<4.1340<4.1985 $\Rightarrow$ 4.0962<4.1985.

Conclusion. Passed.

**Axiom 3: Total comparability (connectedness)**

Statement. $\forall a, b \in X \ \big((a < b) \lor (b < a) \lor (a = b)\big)$.

Test. All pairs are orderable (e.g., 4.0623 vs 4.1340; 4.0538 vs 4.2329).

Conclusion. Passed.

**Axiom 4: No greatest element**

Statement. $\forall a \in X \ \exists b \in X \ (a < b)$.

Data-level test. A maximum exists: 4.2329. For a=4.2329, there is no $b \in X$ with a<b.

Conclusion. Failed (by construction on a finite, clipped set).

**Axiom 5: No least element**

Statement. $\forall a \in X \ \exists b \in X \ (b < a)$.

Data-level test. A minimum exists: 4.0469. For a=4.0469, there is no b\in X with b<a.

Conclusion. Failed (by construction on a finite, clipped set).

**Axiom 6: Density**

Statement. $\forall a, b \in X \ \big(a < b \Rightarrow \exists c \in X \ (a < c < b)\big)$.

Test. Adjacent distinct pairs (e.g., 4.0469<4.0529, 4.0696<4.0962) have no value of X strictly between them.

Explanation. Finite sampling, low variance, clipping to [1,5], and rounding discretize values and prevent strict density.

Conclusion. Failed.

Endpoint failures (A4–A5) and density failure (A6) are expected at the data level for Likert-clipped simulations; model-level unboundedness may be noted separately but does not alter these data-level results.

### 4.2.3 Theme 3: Perceived Complexity & Integration — Axiomatic Analysis

Let

$$X = \{3.7852, 3.7712, 3.5077, 3.8352, 3.3881, 3.5687, 3.2782, 4.0467, 3.6600, 4.1780\}.$$

Sorted (unique): 3.2782 < 3.3881 < 3.5077 < 3.5687 < 3.6600 < 3.7712 < 3.7852 < 3.8352 < 4.0467 < 4.1780.

**Axiom 1: Irreflexivity** ($\neg a < a$)

Statement. $\forall a \in X \; \neg(a < a)$.

Test. For each value (e.g., 3.7852<3.7852), the comparison is False.

Conclusion. Passed.

**Axiom 2: Transitivity**

Statement. $\forall a, b, c \in X \left((a < b \wedge b < c) \Rightarrow a < c\right)$.

Test. 3.5077<3.7712<4.0467 ⇒ 3.5077<4.0467; 3.3881<3.5687<3.7852 ⇒ 3.3881<3.7852.

Conclusion. Passed.

**Axiom 3: Total comparability (connectedness)**

Statement. $\forall \; a, b \in X \left((a < b) \vee (b < a) \vee (a = b)\right)$.

Test. All pairs are orderable (e.g., 3.6600 vs 4.1780; 3.7852 vs 3.7712).

Conclusion. Passed.

**Axiom 4: No greatest element**

Statement. $\forall a \in X \, \exists b \in X \, (a < b)$.

Data-level test (sample of 10). A maximum exists: 4.1780. For a=4.1780, no $b \in X$ satisfies a<b.

Conclusion. Failed (by construction on a finite subset).

**Axiom 5: No least element**

Statement. $\forall a \in X \, \exists b \in X \, (b < a)$.

Data-level test (sample of 10). A minimum exists: 3.2782. For a=3.2782, no $b \in X$ satisfies b<a.

Conclusion. Failed (by construction on a finite subset).

**Axiom 6: Density**

Statement. $\forall a, b \in X \, \big(a < b \Rightarrow \exists c \in X \, (a < c < b)\big)$.

Test. Adjacent distinct pairs in the sorted set (e.g., 3.7712<3.7852, 4.0467<4.1780) have no value of X strictly between them.

Explanation. Finite sampling, clipping to [1,5], and rounding discretize values and prevent strict density.

Conclusion. Failed.

As with Themes 1–2, failures of A4–A5 (endpoints) and A6 (density) are expected at the data level for Likert-clipped simulations; model-level unboundedness of the latent normals may be noted separately but does not alter these data-level results.

**4.3 General Clarification on Axiomatic Validity**

While the Monte Carlo simulation generated 10,000 synthetic student responses per theme, the axiomatic analysis in §4.2 was carried out manually on a representative sample of 10 scores to enable step-by-step, symbolic checks. In finite, clipped datasets of this kind, some axioms are structurally difficult or impossible to satisfy. In particular, because scores are clipped to the [1,5] Likert range, Axiom 4 (no greatest element) and Axiom 5 (no least element) necessarily fail (endpoints exist); and because the observed set is finite and discretized, Axiom 6 (density) also fails (adjacent values have no strict intermediate). These outcomes do not signal logical inconsistency; rather, they reflect the inherent limitations of finite, clipped sampling. To complement the manual check, the next subsection provides a programmatic implementation that evaluates all six DLO axioms across the full 10,000 record dataset for each theme. This scalable validation confirms whether the patterns observed in the sample generalize to the entire simulation output.

**4.4 Formal Code Implementation: Axiomatic Consistency at Scale**

The following Python code programmatically tests the six DLO axioms, irreflexivity, transitivity, total comparability, no greatest, no least, density, for each theme on the full 10,000-sample simulation.

**Python Code: Axiomatic Evaluation across Themes**

```python
import numpy as np, pandas as pd
np.random.seed(42)
themes = {
    "Ease of Use & Learnability": {"mean": 4.1169, "std": 0.2709},
    "System Efficiency & Learning Burden": {"mean": 4.1240, "std": 0.0910},
    "Perceived Complexity & Integration": {"mean": 3.7100, "std": 0.2160}
}
n = 10000
df = pd.DataFrame({k: np.clip(np.random.normal(v["mean"], v["std"], n), 1, 5).round(4) for k, v in themes.items()})
a1 = lambda c: all(not (x < x) for x in c)  # irreflexivity
a2 = lambda c: all((not (a < b and b < c)) or (a < c) for a, b, c in zip(*(sorted(set(c))[i:] for i in range(3))))  # transitivity
a3 = lambda c: all((a < b) or (b < a) or (a == b) for i, a in enumerate(sorted(set(c))) for b in sorted(set(c))[i+1:])  # total comparability
a4 = lambda c: all(any(a < b for b in set(c) if b != a) for a in set(c))  # no greatest element (no endpoints, part 1)
a5 = lambda c: all(any(b < a for b in set(c) if b != a) for a in set(c))  # no least element (no endpoints, part 2)
a6 = lambda c: all(any(a < x < b for x in c) for a, b in zip(*(sorted(set(c))[i:] for i in (0, 1))))  # density
R = {"A1": [], "A2": [], "A3": [], "A4": [], "A5": [], "A6": []}
for col in df.columns:
    c = df[col]
    R["A1"].append(a1(c)); R["A2"].append(a2(c)); R["A3"].append(a3(c))
    R["A4"].append(a4(c)); R["A5"].append(a5(c)); R["A6"].append(a6(c))
print("\nAxiom Evaluation (10,000 students – 6 Axioms × 3 Themes):\n")
print(pd.DataFrame(R, index=df.columns).to_string())
```

Figure 2. Python implementation for evaluating axioms by theme

**Output: Axiom Evaluation Results (10,000 students)**

Axiom Evaluation (10,000 students – 6 Axioms × 3 Themes):

|  | A1 | A2 | A3 | A4 | A5 | A6 |
|---|---|---|---|---|---|---|
| Ease of Use & Learnability | True | True | True | False | False | False |
| System Efficiency & Learning Burden | True | True | True | False | False | False |
| Perceived Complexity & Integration | True | True | True | False | False | False |

Table 2. Axiom evaluation results for 10,000 students across three themes

Each theme, Ease of Use & Learnability, System Efficiency & Learning Burden, and Perceived Complexity & Integration, is tested for the following properties of dense linear order without endpoints (Friedman, 1992, p. 62-63):

1. Irreflexivity: $\forall a \in X \neg(a < a)$.
2. Transitivity (standard form): $\forall a, b, c \in X \big((a < b \land b < c) \Rightarrow a < c\big)$.
3. Total comparability (connectedness): $\forall a, b \in X \big((a < b) \lor (b < a) \lor (a = b)\big)$.
4. No greatest element: $\forall a \in X \, \exists b \in X \, (a < b)$.
5. No least element: $\forall a \in X \, \exists b \in X \, (b < a)$.
6. Density: $\forall a, b \in X \big(a < b \Rightarrow \exists c \in X \, (a < c < b)\big)$.

Whether the formal order-theoretic regularities observed in the 10-value subset extend to the full simulation is assessed through automated checks, thereby allowing the model's logical structure to be clarified. Axioms 4 and 5 are failed by construction due to Likert clipping to [1,5], while Axiom 6 is failed because a finite, rounded sample cannot be dense.

## 5. Philosophical Analysis

### 5.1 Limits of Simulation

While the Monte Carlo simulation confirms several order-theoretic properties at scale, irreflexivity, transitivity, and total comparability, it fails to satisfy the properties that require unboundedness or density. In our implementation, no endpoints (A4-A5) fail because scores are clipped to the Likert interval [1,5], which creates both a minimum and a maximum; density (A6) fails because a finite, rounded sample cannot contain a value strictly between every ordered pair. These are structural consequences of discretization and bounded support, not logical contradictions.

**5.2 Friedman, Dense Linear Order, and the Epistemological Limits of Simulation**

As Michael Friedman (1992) emphasizes, modern axiomatizations include an explicit theory of order. The relevant scheme here is dense linear order without endpoints (DLO), comprising irreflexivity, transitivity, total comparability (connectedness), no greatest and no least elements (A4-A5), and density (A6). In particular, Axiom 6 states that for any two values $a<b$, there exists a c with $a<c<b$. While such a dense, endpoint-free structure is well defined in pure mathematics, it cannot be instantiated by simulation data that are finite, clipped, and discretized. Hence, even though our scores satisfy irreflexivity, transitivity, and comparability, they systematically violate no-endpoint conditions and density. This aligns with Friedman's broader point: empirical models that sample finitely from bounded, discrete representations cannot realize the ideal features of an unbounded, dense continuum (Friedman, 1992, pp. 62–67).

**5.3 Synthetic A Priori Modeling**

The simulation is interpreted as a form of synthetic a priori reasoning: it is grounded in empirical data yet structured by formal conditions of intelligibility such as order, transitivity, and comparability. In this sense, the Monte Carlo procedure functions not merely as a statistical device but as an epistemic instrument for probing the form of cognition that renders such data intelligible.

**5.4 Formal Logic vs. Constructive Intuition**

A distinction is drawn by Friedman (1992) between modern logical axiomatization and Kant's account of constructive intuition. Order on a line can be regimented by dense linear order without endpoints (DLO), irreflexivity, transitivity, total comparability, no greatest, no least, and density, all expressible as first-order formulas about "<". Nevertheless, DLO describes an infinite, endpoint-free, dense structure that cannot be realized by finite, discretized data. Kant's claim is that the very representation of the infinite depends on intuition, not on concepts. As he puts it, "Space is represented as an infinite given magnitude" (Friedman, 1992, p. 64-65), and "a line … drawn to infinity … presupposes a representation of space and time that can only depend on intuition" (Prolegomena §12; quoted in Friedman, 1992, p. 64). On this reading, the idea of denseness is not captured by a formula alone: "The notion of infinite divisibility or denseness … cannot be represented by any such formula as 6" (Friedman, 1992, p. 64).

Accordingly, Euclid has no axiom for density; instead, a uniform constructive procedure is provided, e.g., bisecting any given segment, which can be iterated indefinitely and thereby represents infinite divisibility in practice (Friedman, 1992, p. 65). Friedman links this to a modern perspective: constructive rules function like Skolem functions, replacing existential postulates with explicit constructions of the required instances (Friedman, 1992, p. 65). In this Kantian setting, the usual modern split between pure and applied geometry cannot be drawn: "the only way to represent the theory of linear order 1–6 is to provide, in effect, an interpretation that makes it true" (Friedman, 1992, p. 66). Friedman further clarifies the limits of monadic (syllogistic) logic using Kant's discussion at B40: purely monadic resources cannot force infinitude, with k primitive predicates, one can partition a domain into at most $2^k$ maximally specific classes, so no set of monadic formulas can require an infinite extension (Friedman, 1992, pp. 67–68). This technical observation is aligned with Kant's insistence that concepts, understood monadically, cannot contain actual infinity "in themselves," whereas space is given as infinitely divisible in intuition (Friedman, 1992, pp. 64, 67–69). Against this background, the present simulation's outcomes are unsurprising. At the data level, scores are clipped to [1,5] and rounded, so no-endpoint conditions fail, both minima and maxima occur, and density fails, adjacent distinct values lack guaranteed intermediates. These results mark structural limits of finite, discretized sampling rather than logical inconsistency. In Kantian terms, they exemplify the gap Friedman articulates: logical axioms articulate an ideal dense, endpoint-free order, while intuitive construction supplies the means to realize such properties, something a finite simulation can approximate but not instantiate (Friedman, 1992, pp. 64–66).

### 5.5 Concept vs. Intuition in Representing Infinity

Kant draws a sharp line between conceptual representation and intuitive construction, especially regarding infinity. As Friedman emphasizes, modern logical formulas can state infinitary properties (e.g., density as $\forall x \forall y (x < y \rightarrow \exists z\, (x < z < y))$), but such statements do not construct the infinite manifold in the way Kant requires (Friedman, 1992, pp. 64–66). For Kant, the givenness of the infinite, "Space is represented as an infinite given magnitude", depends on pure intuition, not on concepts alone (Friedman, 1992, p. 64). Hence, density is secured not by an existential axiom but by a uniform constructive procedure, e.g., iterated bisection in Euclidean geometry, that can be carried out indefinitely (Friedman, 1992, p. 65). In this light, the Monte Carlo simulation's failure to realize density is not a defect of logic but a consequence of finite, discretized data. The model thereby marks the boundary between formal description and intuitive synthesis: logic can describe the ideal order, while only intuition can produce its infinite divisibility in act (Friedman, 1992, pp. 66–69).

## 6. Visualization and Kantian Continuity

In order to bridge the empirical logic of simulation with Kantian philosophical interpretation, this section presents a visual model that contrasts discrete empirical data with idealized perceptual continuity.

### 6.1 Simulated and Idealized Representations

A histogram from the Monte Carlo simulation of student success scores was generated, and an idealized continuous proxy, a sine curve, was superimposed over the same [1,5] Likert range, mapping [1,5] to $[0, \pi]$ via $x \mapsto \pi(x - 1)/4$. The histogram exhibits the discrete, finite character of empirical sampling, whereas the sine curve visualizes a smooth, uninterrupted profile evocative of Kantian constructive continuity.

#### 6.1.1 Python Code: Monte Carlo Histogram and Sine Function with Tangents

The following code constructs the visualization, adding tangent lines at key points to signify localized cognitive synthesis:

```python
import numpy as np, matplotlib.pyplot as plt

np.random.seed(42)
scores = np.random.normal(4.1, 0.27, 10000)
x = np.linspace(1, 5, 1000)
f = lambda x: np.sin(np.pi * (x - 1) / 4)
df = lambda x: (np.pi / 4) * np.cos(np.pi * (x - 1) / 4)

plt.figure(figsize=(10, 6))
plt.hist(scores, bins=50, density=True, alpha=0.6, color='skyblue',
         label='Monte Carlo Distribution')
plt.plot(x, f(x), 'darkred', lw=2.2, label=r'Kantian Curve: $f(x)=\sin\left(\frac{\pi(x-1)}{4}\right)$')

for x0, c in zip([2, 3, 4], ['darkgreen', 'darkorange', 'purple']):
    y0, m = f(x0), df(x0)
    plt.plot(x, m*(x - x0) + y0, '--', lw=1.5, color=c, label=f'Tangent at x={x0}')

plt.title("Monte Carlo vs. Kantian Continuity", fontsize=14)
plt.xlabel("Simulated Success Score (Likert 1-5)", fontsize=12)
plt.ylabel("Density / Value", fontsize=12)
plt.legend(); plt.grid(True); plt.tight_layout(); plt.show()
```

*Figure 3 . Code generating a histogram and a sine-based continuity curve with tangents*

Together, the histogram, sine curve, and tangent lines represent the epistemological boundary between logic-based simulation and intuition-based perception. This visual framework sets the stage for the philosophical discussion that follows.

**6.2 Tangent Lines and Temporal Synthesis**

The tangent lines drawn at x = 2, 3, 4 are not merely mathematical markers; they visualize what Kant refers to as productive synthesis in time. Each tangent illustrates a localized rate of change, expressing how perception is dynamically generated, not statically received. At x = 2, the curve is rising, positive slope; at x = 3, the slope is zero, a moment of maximal intensity; and at x = 4, it declines, negative slope. These transitions mirror how the mind actively constructs perceptual structures across time.

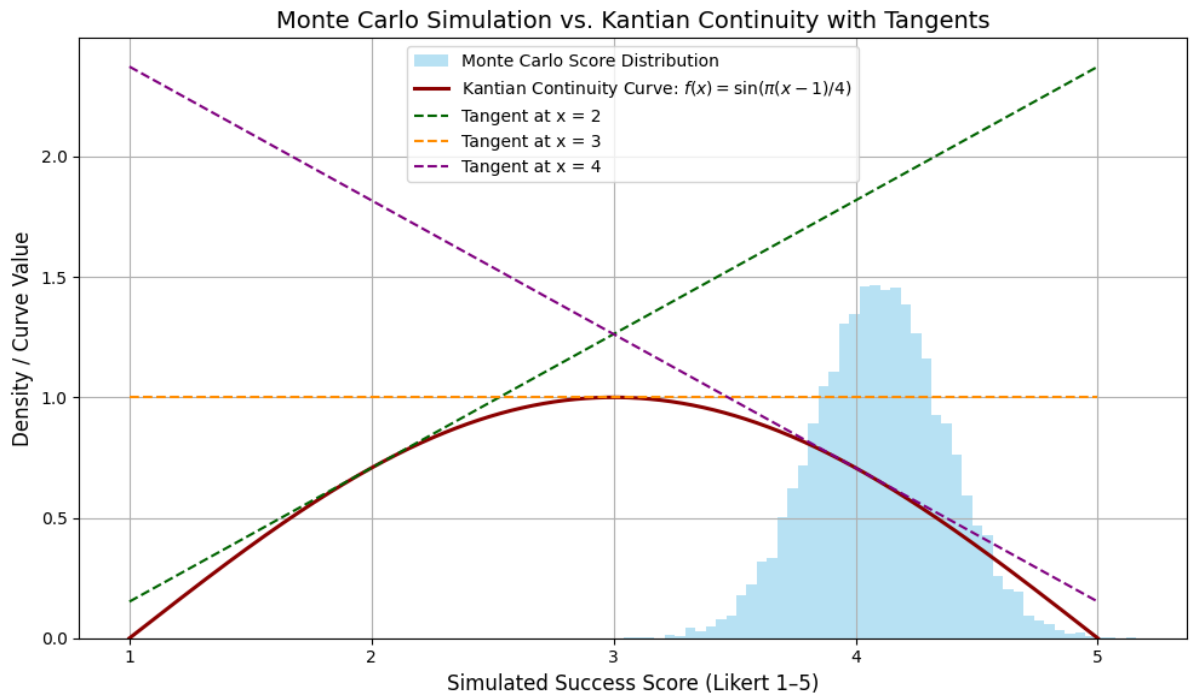

*Figure 4. Monte Carlo histogram of student success scores with a continuous sine curve and tangent lines at three points.*

### 6.3 Histogram vs. Continuum

The histogram, derived from finite Monte Carlo samples, reflects discrete and empirically grounded variation. Its visible gaps and clustered peaks reveal the stochastic nature of simulation-based modeling. In contrast, the sine curve is smooth, continuous, and differentiable at every point, capturing what Kant calls the formal intuition of space. This juxtaposition between histogram and curve illustrates the divide between statistical representation and transcendental ideality.

### 6.4 Philosophical Commentary on Figure 4: Kantian Continuity and the Structure of Simulation

To illustrate the contrast between empirical modeling and Kantian continuity, two representations were plotted on the same figure. The histogram represents simulated success scores for 10,000 synthetic students generated via a Monte Carlo simulation. These scores reflect aggregated Likert-scale values and exhibit a discrete, empirical distribution with observable gaps between values. This pattern captures the randomness and granularity typical of probabilistic modeling processes.

Superimposed on this histogram is a theoretical sine function, defined as $f(x) = \sin\left(\frac{\pi(x-1)}{4}\right)$, which models Kantian continuity across the same [1–5] Likert scale. This function was specifically chosen to map the Likert domain onto a smooth trigonometric curve that rises from 0 to 1 and falls back to 0. The transformation $x \mapsto \frac{\pi(x-1)}{4}$ shifts the domain from [1, 5] to $[0, \pi]$, thus allowing the sine function to express a complete perceptual wave, a rise, a peak, and a decline in perceived success, mirroring how Kantian intuition develops over time. Unlike the histogram, this sine curve is continuous and differentiable at every point, embodying the idea of uninterrupted perceptual flow. To further emphasize this differentiability and to bring out the dynamic epistemic structure of Kantian thought, tangent lines were added at three critical points: x = 2, x = 3, and x = 4. These tangents are computed from the analytical derivative of the sine function,

$$f'(x) = \frac{\pi}{4} \cos\left(\frac{\pi(x-1)}{4}\right),$$

and represent the local linear approximation at each point on the curve. The inclusion of these tangent lines is not merely an aesthetic choice; it is a deeply philosophical one. Without them, the curve might appear smooth, but the inner dynamics, the rates of change and directions of movement, remain visually implicit. With the tangents present, the structure of Kantian continuity can be expressed in its full depth. They make visible what Kant called the productive synthesis of intuition: not merely the coexistence of points on a curve, but the active generation of those points through temporal progression. Each tangent thus functions as a visual embodiment of infinitesimal continuity, a local, determinate moment within a broader, constructed whole. This is crucial because, in Kant's Transcendental Aesthetic, space and time are not given as concepts but as forms of intuition, continuous, infinitely divisible, and constructed through inner sense. The tangent lines visually convey this "constructed continuity" (Friedman, 1992, pp. 58–66). They demonstrate that any point on the curve is also part of a locally linear process, reflecting the idea that perception does not jump between values, but flows in a structured and orderly way. The selected points x = 2, x = 3, and x = 4 were chosen for both their semantic relevance on the Likert scale and their mathematical significance on the sine wave. At x = 2, the slope is positive, indicating a rising phase; at x = 3, the curve peaks with a zero slope; and at x = 4, the slope becomes negative, representing a falling perceptual intensity. These points capture the full spectrum of change, reinforcing the idea of continuous synthesis in time and experience. This entire visualization stands in stark contrast to the empirical variability captured by Monte Carlo simulations. While the histogram reflects student perceptions as discrete data points drawn from a probabilistic process, the Kantian sine curve, especially with its tangent lines, represents an ideal of smooth, uninterrupted, and mathematically structured cognition. The simulated data fails the no-endpoints conditions (A4-A5) and density (A6): A4-A5 are violated because clipping to the [1, 5] Likert bounds introduces least and greatest elements, whereas A6 is violated because a finite, discretized sample need not contain a value strictly between any two distinct scores. This failure, however, does not indicate a contradiction but a limitation imposed by the discrete and finite nature of empirical simulation. The continuous proxy instantiates density on the open interval (1, 5); yet when the domain is taken as [1, 5], endpoints remain, so the no-endpoints conditions (A4-A5) are not satisfied on the closed interval. As such, the figure operates on two levels, empirical and transcendental: the histogram expresses what is measured, while the curve and its tangents express how structured perception might unfold if guided by principles of synthetic a priori reasoning. In sum, this figure is not only a juxtaposition of simulation and theory, it is a philosophical bridge. The empirical histogram shows what the world gives us through measurement; the Kantian curve shows how the mind gives structure to that data through temporal synthesis. The tangent lines, crucially, make this synthesis visible. They transform the curve from a static representation into a dynamic expression of inner time-consciousness. This is why the figure is essential not only for illustrating model results but for articulating the very epistemological foundations upon which the simulation stands (Friedman, 1992, pp. 55–84).

**6.5 Infinite Construction and the Limits of Axiomatics**

As Michael Friedman (1992) argues in his analysis of Euclidean and Hilbertian geometries, the notion of continuity in mathematics concerns the logical forms that structure spatial and temporal constructions. In classical Euclidean practice, continuity is not postulated as an axiom; geometric objects are constructed by finitely repeatable operations without assuming intermediate points. By contrast, modern axiomatizations, e.g., Hilbert's, employ dense linear order without endpoints (DLO), stated in polyadic first-order logic beyond monadic/syllogistic means, and in addition, separately, invoke a second-order completeness principle to strengthen the system (Friedman, 1992, pp. 60–66). Within DLO, Friedman highlights axioms including irreflexivity, transitivity, total comparability, no greatest element (A4), no least element (A5), and density (A6). Axioms 4-5 together state the absence of endpoints; Axiom 6 states that for any two values $a<b$, there exists a third c with $a<c<b$, order-theoretic density.

While such structure is well-defined in pure mathematics, it cannot be fully instantiated by empirical simulations built from discrete samples. In our Monte Carlo data, the basic order axioms (A1-A3) hold, but A4-A5 fail at the data level because clipping to the [1,5] Likert bounds introduces least and greatest elements, and A6 fails because a finite, discretized sample need not contain a value strictly between any two distinct scores. This aligns with Friedman's point that "continuity," understood as infinite density, cannot emerge from discretized approximations alone. As he emphasizes in Kant and the Exact Sciences, the relevant contrast is between formal, symbolic frameworks and the intuitive, constructive acts that unfold within the pure forms of space and time (1992, pp. 60–68). The simulation therefore, marks not a logical inconsistency but an epistemological boundary, a Kantian threshold between symbolic representation and intuitive construction.

**6.6 Euclidean Construction and the Kantian Form of Continuity**

Kant's view of mathematical construction departs from both empirical representation and mere symbolic abstraction. As Friedman explains, Euclidean geometry, understood through its postulates and construction rules, exemplifies an a priori method in which mathematical objects are built by iterative acts rather than inferred from observations. Euclid's "postulates" operate as rules of construction; drawing a straight line between two points or bisecting a segment is an intuitive act unfolding in inner time (Friedman, 1992, pp. 58–60). Features like infinite divisibility, order-theoretic density, are not secured by asserting an existential axiom; rather, they are exhibited by a uniform procedure, iterated bisection, that can be applied indefinitely. In modern logical terms, this behaves like a constructive Skolem function for the ∃ in the density condition (Friedman, 1992, p. 65), whereas axiomatic modern treatments encode density explicitly as

$$\forall a \, \forall b \, (a < b \rightarrow \exists c \, (a < c \wedge c < b)).$$

(Friedman, 1992, pp. 62–63).

By contrast, modern axiomatizations, e.g., Hilbert's, state dense linear order without endpoints in polyadic first-order logic and supplement it with a continuity (completeness) axiom (Friedman, 1992, p. 63). The Kantian-Euclidean stance is mirrored in our visualization: the sine curve with tangents represents "constructed continuity," while the Monte Carlo histogram records finite, discretized outcomes. Thus, density and no-endpoints are realized by construction on an ideal continuum but, as expected, fail in clipped, finite data, precisely the boundary Friedman draws between formal description and intuitive synthesis (Friedman, 1992, pp. 60–66).

**7. Conclusion**

This study builds exclusively upon a prior simulation-based model developed by Kayadibi (2025), which introduced a Monte Carlo framework for modeling student perceptions of generative AI (GenAI) in higher education. That earlier work synthesized Likert-scale survey statistics from a representative dataset to simulate a composite "Success Score" and offered a methodological alternative to conventional meta-analytic techniques. Rather than expanding the dataset or incorporating additional sources, the present study focuses on reinterpreting that model through a Kantian-axiomatic lens to examine its internal logical structure and epistemological scope.

By applying six axioms of dense linear order, ranging from transitivity and comparability to density and the absence of terminal points, the simulation was evaluated not just as a statistical construct but as a cognitive model. At the data level, the basic ordering axioms, irreflexivity, transitivity, and total comparability (A1-A3), were satisfied. By contrast, the no-endpoints conditions, no greatest and no least; A4-A5, and density (A6) failed, as expected, because Likert-bounded, clipped, and finite samples necessarily contain minimum and maximum observed values and need not include a third value strictly between any two distinct scores. These outcomes are thus read not as methodological defects but as indicators of the boundary between finite empirical modeling and the ideal structures articulated in Kantian and modern axiomatic terms. Drawing on Michael Friedman's interpretation of Kant and modern axiomatics, it was argued that the continuity features captured by no-endpoints and density are satisfied in an ideal, constructively continuous setting but cannot be realized in a finite, discretized dataset. In this sense, the "continuum-like" properties belong to a polyadic first-order axiomatization of order (DLO) and to stronger completeness principles beyond DLO, whereas the simulation records only finitely many clipped observations. To illustrate this epistemological divide, a histogram of simulated success scores was contrasted with an idealized sine curve mapped onto the same Likert scale. The histogram, though statistically valid, reflects the discrete and fragmented nature of empirical modeling. The smooth curve, by contrast, serves as a proxy for Kantian constructed continuity, with tangent lines making visible the dynamic, time-based structure of cognition.

Crucially, no new empirical data, instruments, or scenarios were introduced. Instead, an existing model was reframed within a Kantian-axiomatic perspective, showing how formal simulations can reflect, and fail to reflect, the synthetic a priori conditions of human cognition. In this respect, the contribution is an expansion of meaning, not data: a re-positioning of a statistical model as a tool for epistemological inquiry. Future work may adapt this formal-philosophical simulation approach to other domains of AI-assisted learning or investigate designs that more directly engage cognitive structure. Any such advances, however, must continue to recognize the central insight underscored here: the limits of simulation are not merely computational; they are epistemological. No quantitative model, whether simulation, survey analysis, regression, or network modelling, can fully coincide with the pure, intuition-constructed forms that Kant takes to underwrite continuity. Such models are always approximations. Their limits are not merely statistical, sampling error, clipping, rounding, but epistemological: finite, discretized representations cannot realize the endpoint-free density and completeness that belong to ideal structures (Friedman, 1992). Our contribution is therefore not to overcome these limits, but to make them explicit and theoretically legible.

**REFERENCES**

Friedman, M. (1992). Kant and the exact sciences. Harvard University Press.

Kayadibi, S. Y. (2025). Quantifying student success with generative AI: A Monte Carlo simulation informed by systematic review (arXiv Preprint No. arXiv:2507.01062).
https://doi.org/10.48550/arXiv.2507.01062

Veras, M., Dyer, J.-O., Shannon, H., Bogie, B., Ronney, M., Sekhon, H., Rutherford, D., Silva, P., & Kairy, D. (2024). A mixed methods crossover randomized controlled trial exploring the experiences, perceptions, and usability of artificial intelligence (ChatGPT) in health sciences education. *DIGITAL HEALTH*, *10*.
https://doi.org/10.1177/20552076241298485